%% file: main.tex
\documentclass[runningheads]{llncs}
\usepackage{graphicx}
\begin{document}
%
% \title{Structuring Quechua Language using Wikibase}
\title{Getting Quechua Closer to Final Users through Knowledge Graphs}
\author{
Elwin Huaman\inst{1} \and
Jorge Luis Huaman\inst{2} \and
Wendi Huaman\inst{2}
}
\authorrunning{E. Huaman et al.}
\institute{
Semantic Technology Institute (STI) Innsbruck, Austria \\
\email{elwin.huaman@sti2.at} \and 
National University of Altiplano Puno, Peru \\
\email{jorgellhq@gmail.com}, \email{wendi.huamanquispe@gmail.com}
}
\maketitle              % typeset the header of the contribution
\begin{abstract}
Quechua language and Quechua knowledge gather millions of people around the world, especially in several countries in South America. Unfortunately, there are only a few resources available to Quechua communities, and they are mainly stored in PDF format.
In this paper, the Quechua Knowledge Graph is envisioned and generated as an effort to get Quechua closer to the Quechua communities, researchers, and technology developers. Currently, there are 553636 triples stored in the Quechua Knowledge Graph, which is accessible on the Web, retrievable by machines, and curated by users.
To showcase the deployment of the Quechua Knowledge Graph, use cases and future work are described.
\end{abstract}
\section{Introduction}
\label{sec:introduction}
\input{chapter/1-Intro}
\section{Related Work}
\label{sec:related-work}
\input{chapter/2-Related}

\section{Quechua Knowledge Graph}
\label{sec:approach}
\input{chapter/3-Approach}
\section{Feasibility}
\label{sec:feasibility}
\input{chapter/4-Feasibility}
\section{Use Cases}
\label{sec:use-cases}
\input{chapter/5-Use-cases}
\section{Conclusion and Future Work}
\label{sec:conclusion}
\input{chapter/6-Conclusions}
\subsubsection*{Acknowledgements.} We would like to thank: 
David Lindemann, Valeria Caruso, and Ibai Guillen for their fruitful collaboration and technical contribution to this work.
Furthermore, the research leading to these results has been supported by the NexusLinguarum COST Action CA18209
% We want to thank the 4th Summer Datathon on Linguistic Linked Open Data (SD-LLOD-22), which made possible the realization of the Quechua Knowledge Graph.
\bibliographystyle{splncs04}
\bibliography{bib}
\end{document}

%% file: chapter/1-Intro.tex
% WHY
The availability of interoperable linguistics resources is nowadays more urgent in order to save and help under-resourced languages, and their communities. Despite the efforts of the linguistic community, not all languages are represented, nor made accessible in a structured format. For instance, there are only a few resources available for the Quechua language, and they are mostly in PDF format.

% WHAT
In the literature, and to the best of our knowledge, most knowledge bases constructed recently are in well-spread languages and for well-established communities, like English or Spanish, as they take up an overwhelming majority on the Web, while under-resourced languages or indigenous communities receive less attention, for example, there is no structured knowledge graph dedicated to the Quechua community.

% HOW
In order to overcome these limitations, in this paper, we propose the Quechua Knowledge Graph, which aims to support a harmonization process of the Quechua language and knowledge. To do it, we are following a process model for knowledge graph generation~\cite{FenselSAHKPTUW20}, which involves i) knowledge creation, ii) knowledge hosting, iii) knowledge curation, and iv) knowledge deployment phases.
To date, the Quechua Knowledge Graph contains 553636 triples, which are accessible on the Web, retrievable by machines, and curated by the Quechua community.

%STRUCTURE
The remainder of the paper is organized as follows. 
In Section \ref{sec:related-work} the related work is presented, Section \ref{sec:approach} describes the Quechua Knowledge Graph, and the feasibility of it is discussed in Section \ref{sec:feasibility}. Furthermore, in Section \ref{sec:use-cases} we list use cases where the Quechua Knowledge Graph may be used. Finally, we conclude with Section~\ref{sec:conclusion}, providing some remarks and future work plans.

%% file: chapter/2-Related.tex
Knowledge graphs are semantic nets that represent knowledge about certain domains from integrating heterogeneous sources~\cite{FenselSAHKPTUW20}. They have shown to be very useful for applications such as personal assistants (e.g. Alexa), question-answering systems (e.g. WolframAlpha), and search engines (e.g. Google)~\cite{Fensel2020HowToUseKGs}. Content in knowledge graphs can be either open or proprietary, and they can be classified according to those criteria. Open knowledge graphs can be manually curated by humans (e.g., community-driven) or semi-automatically curated based on authoritative knowledge sources. By contrast, proprietary knowledge graphs are restricted to be accessed, therefore the curation process is mostly unknown.
There are various well-known and widely used knowledge graphs that represent general-purpose knowledge, for instance, DBpedia\footnote{\url{https://www.dbpedia.org/}} and Wikidata\footnote{\url{https://wikidata.org/}}. 
%However, to the best of our knowledge, there is no a knowledge graph dedicated to the Quechua language and knowledge. 

\begin{table}[h]
\caption{Comparing approaches built for the Quechua language and knowledge.}
\label{table:quechua-bases}
\begin{tabular}{|p{2.9cm}|p{2.9cm}|p{2.9cm}|p{2.9cm}|}
\hline
\textbf{Description} & \textbf{Wiktionary}              & \textbf{Wikidata}                   & \textbf{Qichwabase}                 \\ \hline
RDF support                 & No                & Yes                   & Yes                   \\ \hline
Reuse of ontologies         & No                & Yes                   & Yes                   \\ \hline
Scalability                 & Wiktionary server & Wikibase              & Wikibase              \\ \hline
Learning curve       & User should learn code templates & User selects classes and properties & User selects classes and properties \\ \hline
Anyone can edit             & Yes               & Yes                   & Yes                   \\ \hline
Search                      & Yes               & Yes                   & Yes                   \\ \hline
Track changes               & Yes               & Yes                   & Yes                   \\ \hline
Community-driven            & Yes               & Yes                   & Yes                   \\ \hline
Fixed structure of entries  & No                & Yes                   & Yes                   \\ \hline
Automatic (Bots) & Yes, restricted   & Yes, restricted       & Yes                   \\ \hline
Use-oriented                & Human         & Human, machine & Human, machine \\ \hline
Covers                      & Quechua language         & Quechua language and knowledge & Quechua language and knowledge \\ \hline
\end{tabular}
\end{table}

In the context of the Quechua language and knowledge, to the best of our knowledge, there is no dedicated knowledge graph for it. The most approximate knowledge graph to be considered on this matter may be Wikidata, which represents multilingual knowledge in a structured format. Furthermore, it is important to mention Wiktionary\footnote{\url{https://qu.wiktionary.org/}} as a database for the Quechua language. We compare those approaches in Table~\ref{table:quechua-bases}.

%% file: chapter/3-Approach.tex
In this section, we describe the workflow followed to build the Quechua Knowledge Graph, which is available at \url{https://qichwa.wikibase.cloud/}. This includes the creation, hosting, curation, and deployment phases.
\begin{figure}
    \centering
    \includegraphics[width=0.98\textwidth]{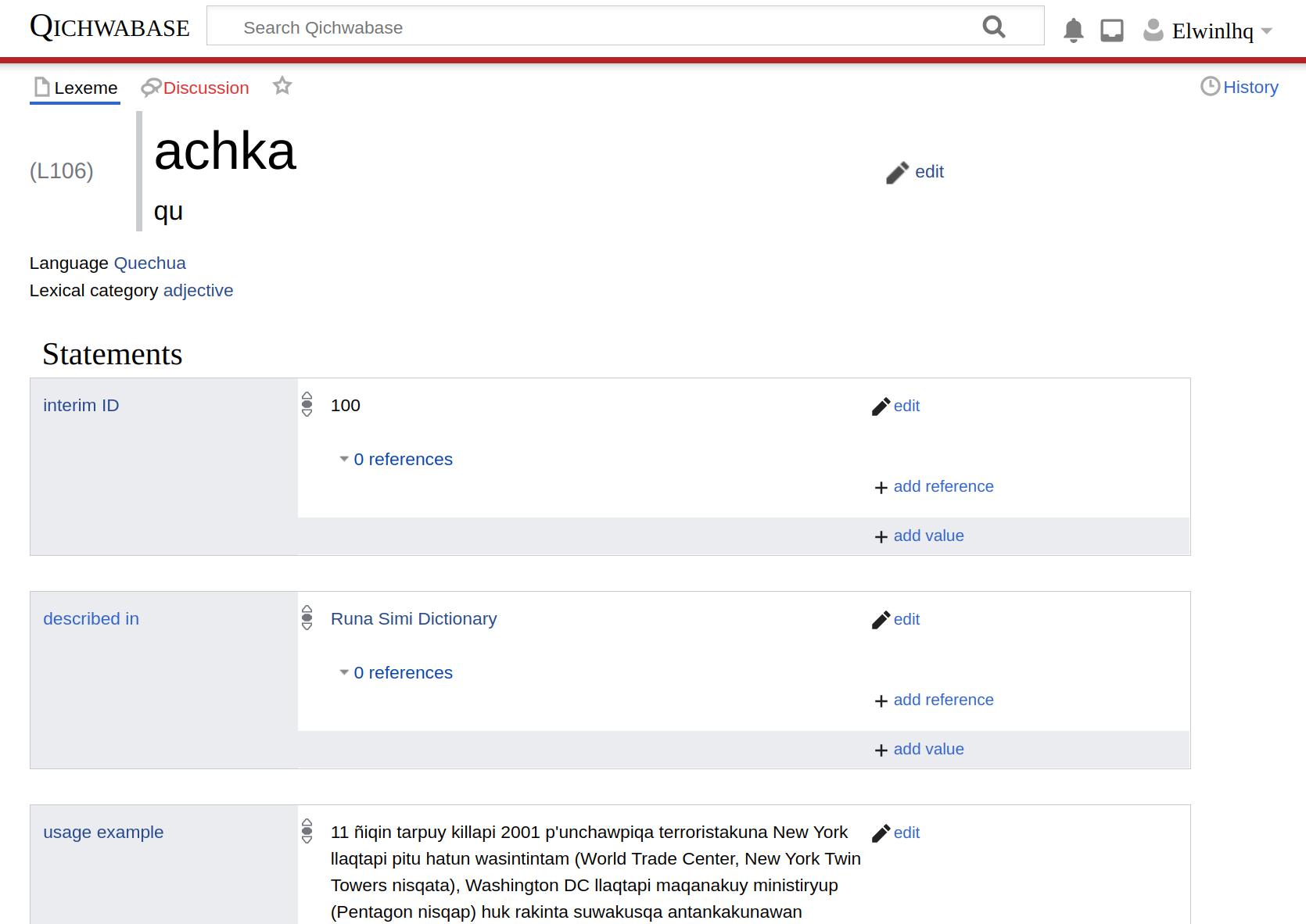}
    \caption{Quechua Knowledge Graph interface (I).}
    \label{fig:QKG1}
\end{figure}
\subsection{Knowledge Creation}
\label{subsec:kcreation}
\begin{figure}
    \centering
    \includegraphics[width=0.98\textwidth]{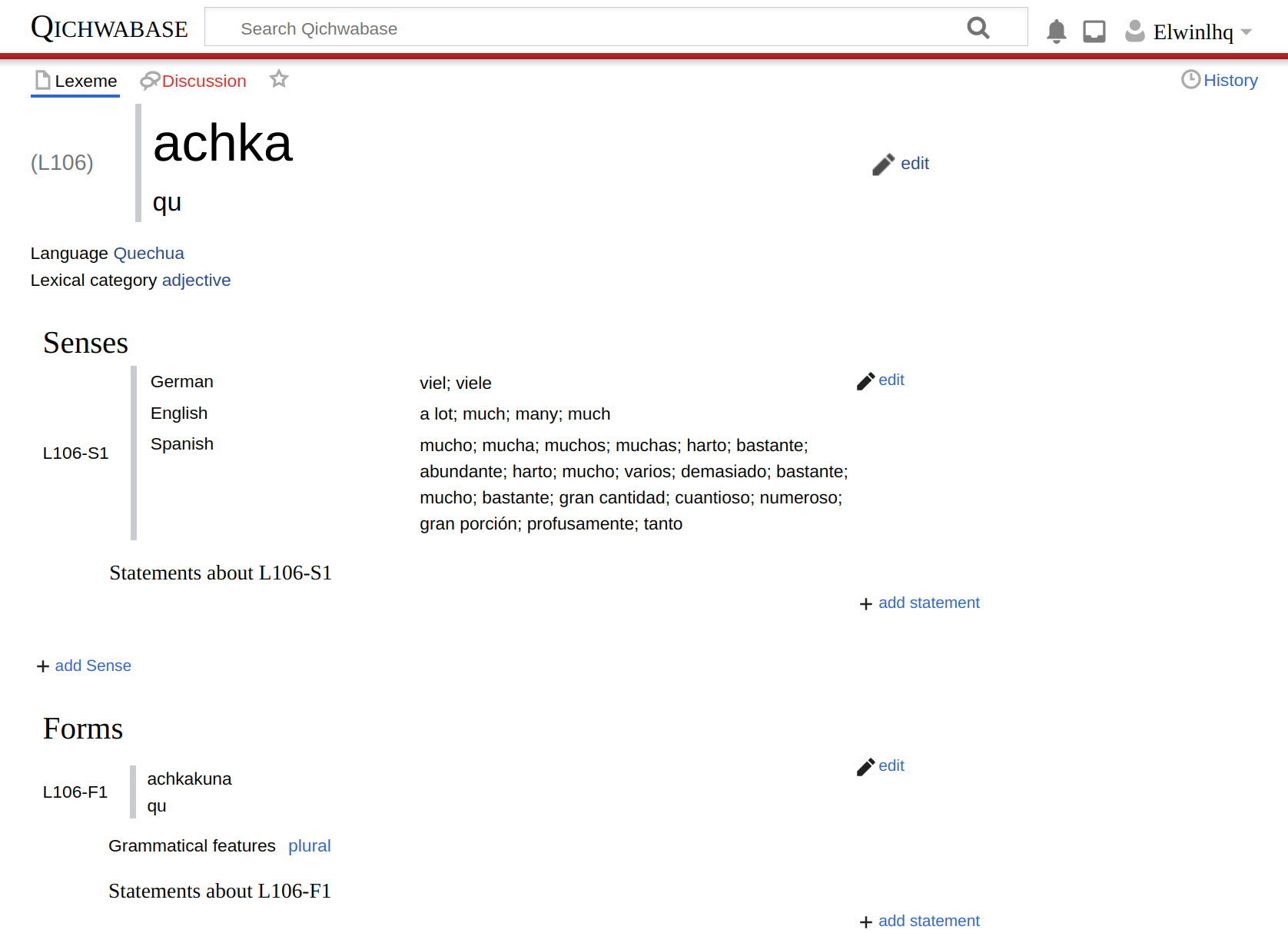}
    \caption{Quechua Knowledge Graph interface (II).}
    \label{fig:QKG2}
\end{figure}

It describes the process of acquiring different sources, modelling them, and applying the models to incoming knowledge. It can be described as follows:
\begin{itemize}
    \item \textbf{Identifying sources}. In the Quechua Knowledge Graph, we first identify sources, like Quechua dictionaries, vocabularies, and so on. One of the sources we started to work with was the Runasimi Dictionary\footnote{\url{https://runasimi.de/}}, which contains 22866 Quechua words described in different languages.
    \item \textbf{Defining domain specifications}. Domain specifications model what properties or features must be described for each type of instance in the Quechua Knowledge Graph, for instance, a lexeme type should be described with language category (e.g., Quechua), lexical category (e.g., noun), described in (e.g., the source it comes from), usage example (e.g., text and a reference link where the text comes from), senses, forms. See Fig. \ref{fig:QKG1} and Fig.~\ref{fig:QKG2}.
\end{itemize}
\subsection{Knowledge Hosting}
The Wikibase Cloud\footnote{\url{https://www.wikibase.cloud/}} hosts the Quechua Knowledge Graph we build. Furthermore, the Wikibase infrastructure provides various components, for instance, some of them are:
\begin{itemize}
    \item \textbf{MediaWiki}, it allows installing extensions to customize Wikibase instances.
    \item \textbf{Wikibase}, allows knowledge to be represented as a structured data repository. For instance, Wikibase provides tools like the  WikibaseIntegrator\footnote{\url{https://github.com/LeMyst/WikibaseIntegrator}}, which is a Python library used for automatizing tasks or creating bots for inserting or editing data.
    \item \textbf{MariaDB\footnote{\url{https://mariadb.com/}} database} that allows data, user, and permission management.
    \item \textbf{SPARQL\footnote{\url{https://www.w3.org/TR/sparql11-overview/}} endpoint} for querying and exploiting the knowledge.
\end{itemize} 
Entering data is done by a bot, which inserts data into the Quechua Knowledge Graph based on identified sources and defined models (see Section~\ref{subsec:kcreation}). Currently, there are 553636 triples contained in the Quechua Knowledge Graph.
\subsection{Knowledge Curation}
To build a high-quality Quechua Knowledge Graph, it is important to curate the knowledge contained in it. A manual curation is not recommended due to the large size of knowledge graphs. Therefore, there are some tools that can automatize the curation tasks and can be used or configured in the Quechua Knowledge Graph. Those tools are:
\begin{itemize}
    \item \textbf{OpenRefine}\footnote{\url{https://openrefine.org/}}, it is very useful for cleaning and normalizing data, furthermore, it can also handle different formats, as well as, it provides an entity linking feature between the Quechua Knowledge Graph and Wikidata.
    \item \textbf{WikibaseImport}\footnote{\url{https://github.com/Wikidata/WikibaseImport}}, it allows importing data from Wikidata into the Quechua Knowledge Graph.
    \item \textbf{EntitySchema}\footnote{\url{https://www.mediawiki.org/wiki/Extension:EntitySchema}}, it allows storing domain specifications (in the form of Shape Expression\footnote{\url{https://shex.io/}} Schemas) on wiki pages, and use them for knowledge valildation.
\end{itemize}
\subsection{Knowledge Deployment}
\begin{figure}
    \centering
    \includegraphics[width=0.98\textwidth]{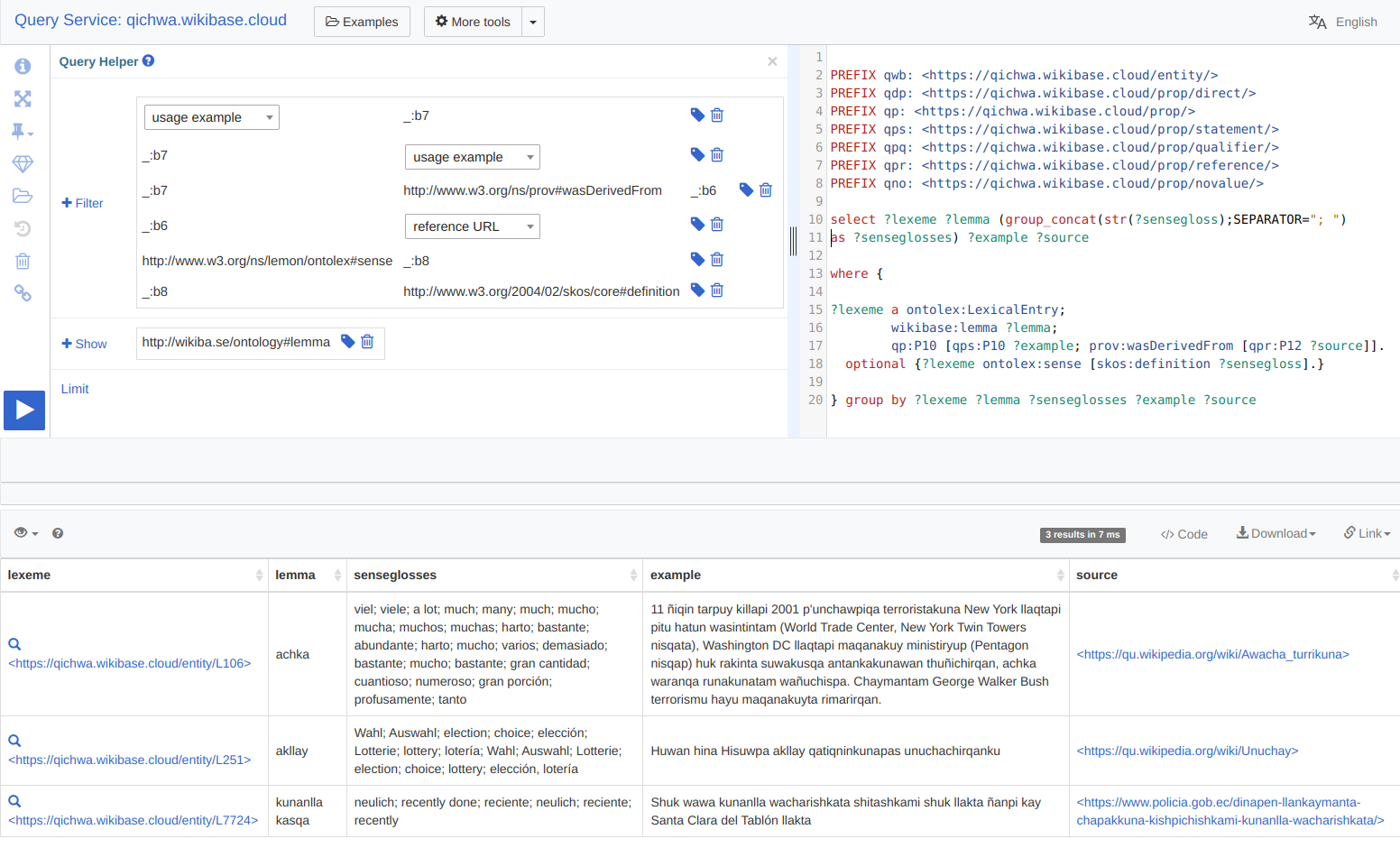}
    \caption{Query displaying lexemes that have multilingual sense descriptions, usage examples, and their source references.}
    \label{fig:QKG3}
\end{figure}

We have built the Quechua Knowledge Graph, which provides various ways of being easily consumable or deployed:
\begin{itemize}
    \item \textbf{An Enhanced GUI}, which allows users to edit faster and more productive, for instance, by adding and editing statements in the form of triples into the Quechua Knowledge Graph, as shown above in Fig.~\ref{fig:QKG1} and Fig.~\ref{fig:QKG2}.
    \item \textbf{A SPARQL endpoint}, it is the standard query service of the Quechua Knowledge Graph, and it is available at \url{https://qichwa.wikibase.cloud/query/}. A screenshot of a query\footnote{The query can be executed here: \url{https://tinyurl.com/2laryjez}} is displayed in Fig.~\ref{fig:QKG3}.
    \item \textbf{Exporting}, the knowledge contained in the Quechua Knowledge Graph can be queried and then exported in various formats (JSON, CSV, HTML, ...).
\end{itemize}

%% file: chapter/4-Feasibility.tex
In order to validate the feasibility of the Quechua Knowledge Graph, we analyse it from a technological, social, and organizational point of view.
\subsection{Technological risks}
\label{subsec:tech-risks}
There are several factors that may compromise the success of the Quechua Knowledge Graph. First of all, is likely to fail without dedicated tools to support tasks, such as i) schema modelling, ii) population of modelled schemas, iii) Knowledge assessment, and iv) knowledge querying. Currently, what is used in the Quechua Knowledge Graph are the built-in features of Wikibase\footnote{\url{https://wikiba.se/}}, which allows one to specify the classes, properties, and constraints at the terminological level (TBox). Furthermore, the initialization and completion of those tasks are done by the community. For instance, defining classes for describing linguistic concepts (verbs, nouns, etc.).
Similarly, tools are needed for the curation\cite{HuamanF21} of the instance level (ABox) of the Quechua Knowledge Graph, which meanly addresses the assessment, cleaning, and enrichment of the knowledge graph. For instance, evaluating the correctness and completeness is not a straightforward task. Currently, Wikibase integrates OpenRefine\footnote{\url{https://openrefine.org/}} for data cleaning and reconciliation.
Furthermore, it is necessary to scale up the tools so that they can handle a large Quechua Knowledge Graph.
\begin{figure}
    \centering
    \includegraphics[width=0.98\textwidth]{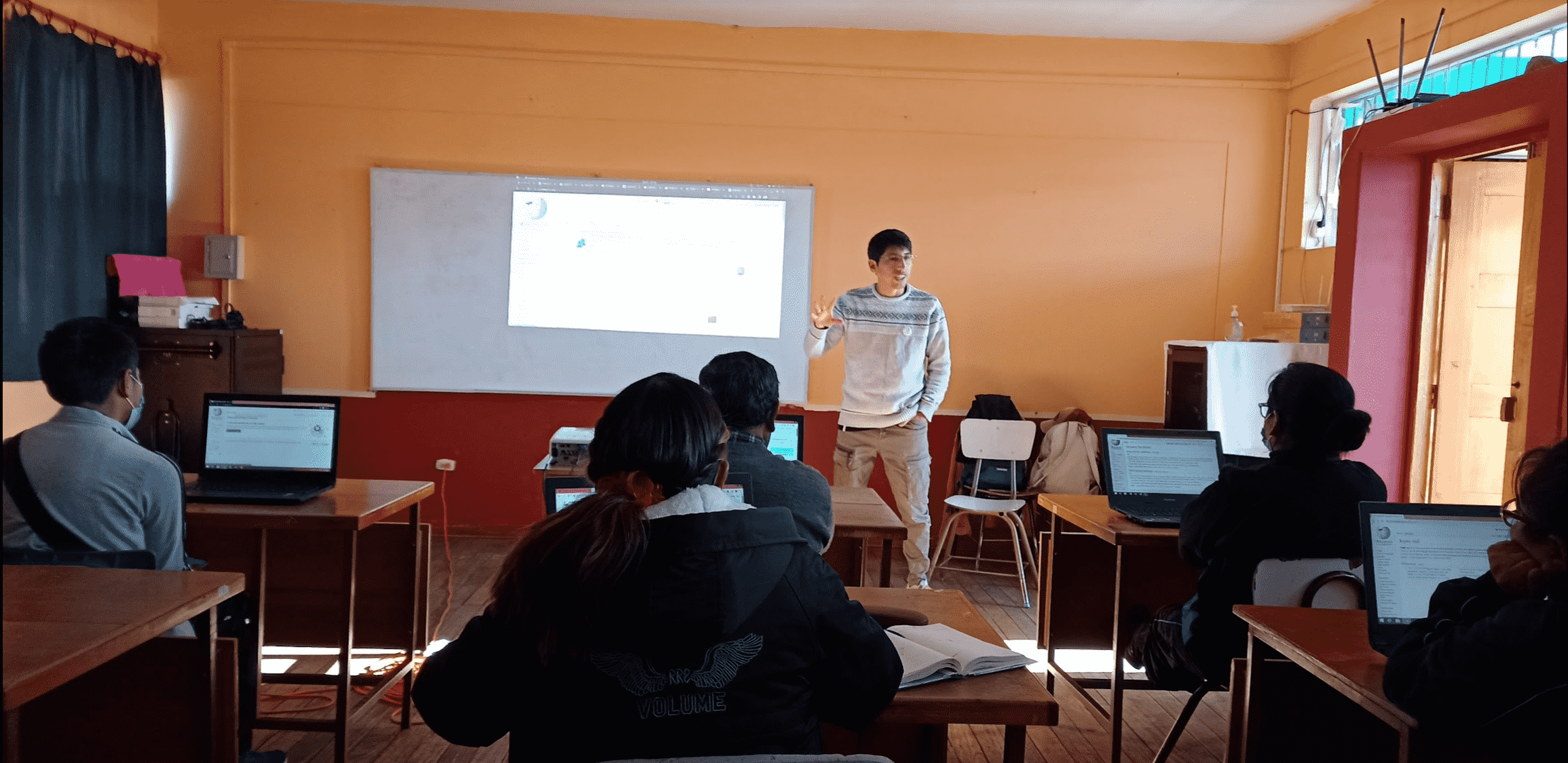}
    \caption{Quechua community event in Nuñoa (Peru), where the Quechua Knowledge Graph is presented and discussed.}
    \label{fig:QKG4}
\end{figure}
\subsection{Social and Organizational risks}
\label{subsec:social-risks}
From a social and organizational point of view, there is one main factor that may endanger the success of the Quechua Knowledge Graph. Without a Quechua community, the initiative would certainly fail. However, participation in the Quechua Knowledge Graph is a rewarding and self-promoting activity. It means that Quechua speakers and users are better off if they participate and contribute so i) a high quality knowledge base can be built and used for developing applications, and ii) a collaborative mentality can be achieved throughout the community, see Fig.~\ref{fig:QKG4}.

%% file: chapter/5-Use-cases.tex
The Quechua Knowledge Graph, as described in Section \ref{sec:approach}, aims to get the Quechua language and knowledge closer to end-users. For that, a knowledge graph generation lifecycle has been followed. In order to leverage the usefulness of the Quechua Knowledge Graph, we list some use cases where it can be used:

\begin{itemize}
    \item \textbf{Question Answering.} The Quechua Knowledge Graph may be used as an interface to answer simple queries, which are equivalent to triples (Subject, Predicate, Object), e.g., (Peru, capital city, Lima). However, complex queries would need different approaches.
    \item \textbf{Dialogue Systems.} They provide a more natural and friendly interaction with final users than question answering systems. In this context, a dialogue system can take advantage of the explicit semantics declared in the Quechua Knowledge Graph for performing reasoning and creating language models.
    \item \textbf{Entity linking.} It allows identifying the same resource across various knowledge bases. For instance, the entities contained in the Quechua Knowledge Graph might be linked to Wikidata and Wikipedia articles, so the representation of entities is richer.
    \item \textbf{Knowledge Validation.} It is a critical task that verifies statements (or facts) against a knowledge base~\cite{HuamanKF20}, in this context the Quechua Knowledge Graph aims to represent knowledge about persons, organizations, places, publications, etc., so the Quechua Knowledge Graph can be used for validating or supporting the semantic correctness of statements.
    \item \textbf{Collaborative Community.} In scattered communities, the mentality might be competitive rather than collaborative~\cite{BenjaminsFG98}, e.g., ``if another Quechua variant wins, then my Quechua variant loses.'' Or ``if I make my dataset available to others, then others will profit from my dataset, and I risk that they may outperform me.'' However, the Quechua Knowledge Graph allows everyone to contribute and make fruitful discussions about the representation of an entity (e.g., Places, Words, etc.) and the historical discussions and arguments are saved and made accessible to anyone.
\end{itemize}

There are more possible use cases where the Quechua Knowledge Graph is applicable. Here, we presented some of them to give an idea about how useful and necessary it is to have a Knowledge Graph for the Quechua community.

%% file: chapter/6-Conclusions.tex
In this paper, we have described our effort for building the Quechua Knowledge Graph from scratch. First, we identified relevant sources and defined models (or domain specifications). Second, we set up a Wikibase instance and programmed a bot for ingesting data. Then, we take advantage of Wikibase features for addressing curation tasks. Finally, we provide various ways of consuming the stored knowledge. Moreover, we also validate the Quechua Knowledge Graph by presenting it to the Quechua community, and by analysing its technological, social, and organizational feasibility. Last but not least, we listed use cases where the Quechua Knowledge Graph can be used.

The limitations of our approach are for example the dependence of the Quechua Knowledge Graph on external services and features. However, Wikibase offers various services that might be very convenient for deploying the Quechua Knowledge Graph. Finding sources in the Quechua language is still a complex process, but we are building and engaging the Quechua community to help to tackle this limitation.

As a next step, we will define more domain specifications (e.g., person and place domains) that can be used for ingesting knowledge about various domains. Furthermore, we will develop applications (e.g., chatbots) powered by the Quechua Knowledge Graph.

%% file: main.bbl
\begin{thebibliography}{1}
\providecommand{\url}[1]{\texttt{#1}}
\providecommand{\urlprefix}{URL }
\providecommand{\doi}[1]{https://doi.org/#1}

\bibitem{BenjaminsFG98}
Benjamins, V.R., Fensel, D., G{\'{o}}mez{-}P{\'{e}}rez, A.: Knowledge
  management through ontologies. In: Reimer, U. (ed.) {PAKM} 98, Practical
  Aspects of Knowledge Management, Proceedings of the Second International
  Conference, Basel, Switzerland, October 29-30, 1998. {CEUR} Workshop
  Proceedings, vol.~13. CEUR-WS.org (1998),
  \url{http://ceur-ws.org/Vol-13/paper5.ps}

\bibitem{Fensel2020HowToUseKGs}
Fensel, D., {{S}}im{{s}}ek, U., Angele, K., Huaman, E., K{\"a}rle, E.,
  Panasiuk, O., Toma, I., Umbrich, J., Wahler, A.: How to Use a Knowledge
  Graph, pp. 69--93. Springer International Publishing, Cham (2020).
  \doi{10.1007/978-3-030-37439-6\_3},
  \url{https://doi.org/10.1007/978-3-030-37439-6\_3}

\bibitem{FenselSAHKPTUW20}
Fensel, D., Simsek, U., Angele, K., Huaman, E., K{\"{a}}rle, E., Panasiuk, O.,
  Toma, I., Umbrich, J., Wahler, A.: Knowledge Graphs - Methodology, Tools and
  Selected Use Cases. Springer (2020). \doi{10.1007/978-3-030-37439-6},
  \url{https://doi.org/10.1007/978-3-030-37439-6}

\bibitem{HuamanF21}
Huaman, E., Fensel, D.: Knowledge graph curation: {A} practical framework. In:
  IJCKG'21: The 10th International Joint Conference on Knowledge Graphs,
  Virtual Event, Thailand, December 6 - 8, 2021. pp. 166--171. {ACM} (2021).
  \doi{10.1145/3502223.3502247}, \url{https://doi.org/10.1145/3502223.3502247}

\bibitem{HuamanKF20}
Huaman, E., K{\"{a}}rle, E., Fensel, D.: Knowledge graph validation. CoRR
  \textbf{abs/2005.01389} (2020), \url{https://arxiv.org/abs/2005.01389}

\end{thebibliography}
